\documentclass[twocolumn,aps,amsmath,amssymb,longbibliography,pr]{revtex4-1}
\usepackage{bm}
\usepackage{epsfig}
\usepackage{amsmath}
\usepackage{color}
\newcommand{\nix}[1]{}
\usepackage[unicode=true,colorlinks=true,urlcolor=blue,citecolor=blue]{hyperref}
\usepackage{ulem}

\begin{document}

\title{Weak Localization in Systems with Chiral Spin Textures and Skyrmion Crystals
}

\author{K.~S.~Denisov$^{1,2}$}
\email{denisokonstantin@gmail.com} 
\author{L.~E.~Golub$^{1}$}
\affiliation{$^{1}$Ioffe Institute, 194021 St.~Petersburg, Russia}
\affiliation{$^{2}$Lappeenranta-Lahti University of Technology, FI-53851 Lappeenranta, Finland}

\begin{abstract}
Theory of interference-induced quantum corrections to conductivity is developed for two-dimensional systems with chiral spin textures including skyrmions. 
The effect of exchange interaction between electrons and spin textures on 
weak localization of electronic waves is studied.
The spin dephasing rates 
are calculated as functions of the spin texture size.
The anomalous magnetoresistance is shown to be governed by 
the  size and magnetization spatial distribution of the spin textures. 
The effect of average magnetization-induced spin splitting on weak localization is analyzed. 
The sign-alternating weak-antilocalization magnetoresistance is demonstrated for skyrmion crystals.  
We argue that analysis of the low-field magnetoresistance serves as an independent tool for an experimental detection of chiral spin textures and, in particular, skyrmions.
\end{abstract}

\maketitle


\textit{Introduction.}
The ongoing decade has been featured by an impressive rise of chiral magnetism. 
Various chiral spin textures, such as 
magnetic skyrmions~\cite{wiesendanger2016nanoscale, fert2017magnetic, soumyanarayanan2017tunable, pollard2017observation, Lin-SkyrmionObservation}
and 
antiskyrmions~\cite{Parkin,hoffmann2017antiskyrmions}, 
skyrmion lattices~\cite{muhlbauer2009skyrmion,B20_FeCoGe},
merons and bimerons ~\cite{yu2018transformation, gobel2018magnetic, kharkov2017bound}
are being extensively studied both experimentally and theoretically. 
The interest is especially heated by the fact that 
the chiral nature of such textures opens up a novel physics 
with a number of intriguing phenomena, such as 
the topological Hall  effect~\cite{Tatara,SpinGlass2,BrunoDugaev,raju2017evolution,Denisov2018PRB},
its reciprocal analog -- the skyrmion Hall effect~\cite{everschor2014real,jiang2017direct}, the anomalous Nernst effect~\cite{Behnia2017} or nontrivial magnon-skyrmion interaction~\cite{Magnon-Skyrmion}. 
Moreover, it is believed that a chiral spin order would manifest itself in a whole variety of solid-state phenomena related to spin interactions.

The electric transport in media with spin skyrmions or other chiral spin textures is modified due to an exchange interaction between itinerant carriers and localized spins forming such textures. 
A carrier propagating in space and interacting with spin textures experiences rotations of its spin. 
In particular, when a texture size is comparable with the electron de-Broigle wavelength, the interaction with textures 
results in spin-dependent electron scattering
~\cite{prl_skyrmion, ishizuka2018spin}. 
Scattering by chiral spatial pattern
affects the electron transport in a nontrivial way. 
For instance, the interference between single and double electron scattering events inside one skyrmion generates the transverse electron flux leading to the Hall response~\cite{Tatara,prl_skyrmion,ishizuka2018spin,nakazawa2018weak}.

Spin-dependent scattering is known to affect electron transport in metallic systems due to the weak localization effect~\cite{Rammer_quant_transp,HLN}. Anomalous magnetoresistance in classically low magnetic fields is caused by magnetoinduced breaking of interference of electron waves passing scattering paths related by the time-inversion. The spin dependence of the scattering amplitude results in dramatic changes of the quantum conductivity correction up to the change of its sign, the phenomenon also known as weak antilocalization.
The anomalous magnetoresistance is enhanced in systems with not extremely large product  $k_\text{F}\ell$ where $k_\text{F}$ is the Fermi wavevector and $\ell$ is the mean free path. 
In systems with chiral spin textures $\ell$ typically does not exceed tens of nanometers~\cite{zang2011dynamics,elias2017steady}, which suggests 
the importance of the interference corrections.

In two-dimensional (2D) systems, the weak localization to antilocalization transitions are studied mainly in semiconductor heterostructures with spin-orbit splitting of the energy spectrum, for review see Ref.~\cite{SST_GlazovGolub_review}. 
Investigations of the spin-flip scattering effect on weak localization are performed for structure-asymmetric $n$-type non-magnetic and magnetic heterostructures~\cite{RomanovAverkiev,PorubaevGolub_110,GaMnAs_MacDonald} and for systems 
with spin-orbit scattering~\cite{Tellur,SSC,FTP,SO-scattering,PorubaevGolub_holes}.
In this work we demonstrate that the low-field magnetoresistance in systems with spin textures is featured by a specific behavior that cannot be attributed to neither spin-orbit nor magnetic-impurity scattering,
thus, its experimental observation would unambiguously indicate the presence of chiral spin order in a system. 

\begin{figure}
	\centering	
	\includegraphics[scale=0.6]{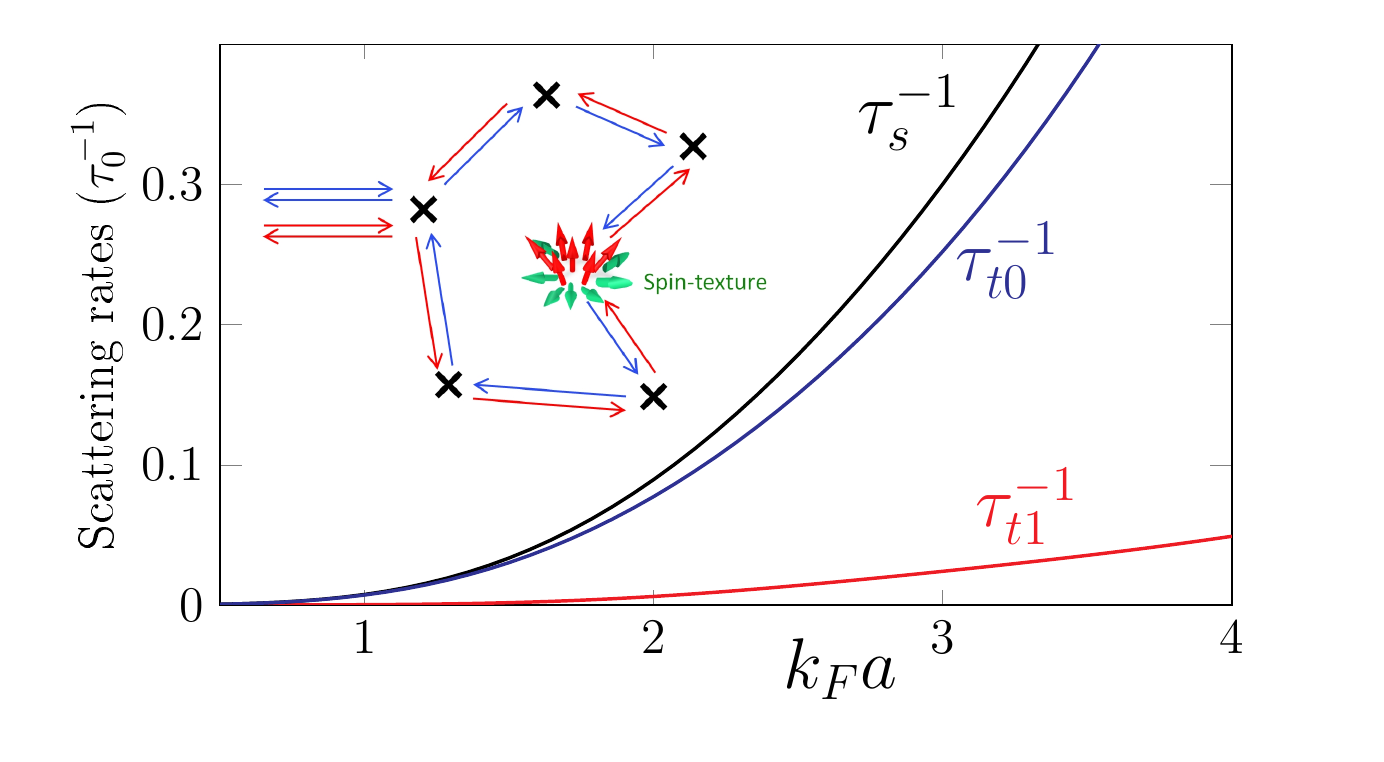}
	\caption{The dependence of the scattering rates for singlet~($\tau_{s}^{-1}$) and triplet~($\tau_{t0,t1}^{-1}$) channels on a spin texture radius 
	for the spin spatial distribution Eq.~\eqref{eq_prof1} with $\alpha = \pi/2$. Inset shows two time-inverted electron trajectories with scattering by both impurities and a spin texture.}
	\label{f1}
\end{figure}

In this Letter, we theoretically investigate the weak localization effect in two-dimensional systems with chiral spin textures. 
We start with description of weak localization 
for the disordered array of spin textures, when each texture causes an additional carrier scattering, see inset to Fig.~\ref{f1}. 
Then we investigate the case of a skyrmion crystal, assuming that skyrmions form a 
regular lattice. 
We demonstrate that the chiral spin pattern in the real space 
affects the magnetoresistivity differently, depending upon its spatial size and the inner structure. Our results suggest that the presence of chiral spin textures can be experimentally detected analyzing 
low-field magnetoresistance.


\textit{Disordered array of chiral spin textures.}
We consider a 2D layer containing randomly distributed chiral spin textures with the spin profile $\boldsymbol{S}(\boldsymbol{r})$ of the following shape: 
\begin{equation}
\boldsymbol{S}(\bm r) = (\bm S_{\parallel}(r,\phi), S_{z}(r)),
\quad S_x \pm iS_y = S_{\parallel}(r)\text{e}^{\pm i\left(\chi \phi + \gamma\right)},
\end{equation}
where $S_z,S_{\parallel}$ are the spin texture out-of-plane and in-plane components, respectively. 
The textures are featured by 
the in-plane rotation of spin, 
the parameters describing this rotation ($\chi$ and phase $\gamma$) 
are determined by a microscopic mechanism behind its formation~\cite{NagaosaNature}.
For instance, when a chiral spin pattern appears due to a spin-orbit interaction, its rotation is given by 
$|\chi|=1$ for linear and $|\chi|=3$ for cubic in momentum spin-orbit splitting of a carrier spectrum~\cite{denisov2018hall}.

We assume that the magnetization contains a homogeneous part $S_0 \bm{e}_z$ normal to the 2D plane and the deviation of magnetization $\delta \bm S(\bm r) = \bm S(\bm r) - S_0 \bm{e}_z$.
%
The Hamiltonian of the electron exchange interaction with the spin texture is given by
\begin{equation}
\mathcal{H}= - h_{ex} S_0 \hat{\sigma}_z-\sum_i h_{ex} 
\hat{\bm \sigma} \cdot \delta \bm S(\boldsymbol{r}-\boldsymbol{R}_i),
\end{equation}
where $h_{ex}$ is the exchange interaction constant, $\hat{\bm \sigma}$ is the vector of Pauli matrices, and the sum runs over spin textures located 
in random points~$\boldsymbol{R}_i$.

In our model, 2D electron band structure 
consists of two parabolas shifted by $\Delta \equiv 2|h_{ex} S_0|$, so
the spin-up and spin-down electrons at the Fermi level have different Fermi wavevectors $k_{\text{F}1,2}=\sqrt{2M(E_\text{F}\pm \Delta/2)}/\hbar$ with $E_\text{F}$ and $M$ being the Fermi energy and electron effective mass, respectively. 
%
In what follows we assume that the mean free path $\ell_{1,2}$ within each spin subband is determined by electron scattering by nonmagnetic impurities.
There are two distinct regimes of weak localization depending on the relation between $\Delta$ and $\ell_{1,2}$. If  the difference $|k_{\text{F}1}-k_{\text{F}2}|$ is much smaller than the inverse scattering lengths $1/\ell_{1,2}$ then the coherence between the spin-splitted subbands is important, and one can neglect the presence of spin splitting $\Delta$. In the opposite case $|k_{\text{F}1}-k_{\text{F}2}| \gg \ell_{1,2}^{-1}$ the coherence is preserved only inside the subbands, and the spin-flip scattering acts as additional dephasing. 
In this paper we consider both these regimes.

%


We start with 
the weak localization effect for the coherent subbands regime $|k_{\text{F}1}-k_{\text{F}2}| \ll \ell_{1,2}^{-1}$. The conductivity correction for 2D systems in a weak perpendicular magnetic field $\bm B$ is given by~\cite{HLN}
\begin{equation}
\label{d_sigma}
\Delta\sigma(B) = {e^2\over 2\pi h} \left[2f_2\left({B\over B_{t1}}\right)+f_2\left({B\over B_{t0}}\right)-f_2\left({B\over B_s}\right) \right],
\end{equation}
where $f_2(x)=\psi(1/2+1/x)+\ln{x}$ with $\psi$ being the digamma function. 
The characteristic magnetic fields for singlet ($B_s$) and triplet states with the angular momentum projection equal to 0 and 1 ($B_{t0,t1}$) are as follows
\begin{equation}
B_i = B_\phi \left( 1 + {\tau_\phi\over\tau_i} \right),
\qquad i=t1,t0,s,
\end{equation}
where 
$\tau_\phi$ is the dephasing time, and the characteristic magnetic field 
$B_\phi=\hbar\tau/(2|e|\ell^2\tau_\phi)$, 
where $\ell$ and $\tau$ are the mean free path and transport scattering time in the subbands.
%
Additional dephasing described by the three times $\tau_{t1,t0,s}$ arises due to spin-dependent scattering
by spin textures.


The amplitude of electron elastic scattering in a system with the textures and ordinary impurities has the form
\begin{equation}
\hat{V}_{\bm k \bm k'} 
= V_0 \hat{I} \sum_j \text{e}^{i\bm q\cdot \bm \rho_j} 
- h_{ex}k_F^{-2}
\hat{\bm \sigma}\cdot {\bm m}(\bm q) \sum_i \text{e}^{i\bm q\cdot \bm R_i},
\end{equation}
where $\bm q = \bm k-\bm k'$, $\hat{I}$ is the $2\times 2$ unit matrix, and $V_0$ is the amplitude of spin-independent scattering by impurities
located in the points $\bm \rho_j$.
The function $\bm m(\bm q)$ is the Fourier transform of a spin texture $\delta \bm S(\bm r)$: 
\begin{align}
\label{m_par}
& m_x(\bm q) \pm i m_y(\bm q) = -i \text{e}^{\mp i (\chi \varphi_{\bm q}+\gamma)} m_\parallel(q),\\
& m_{z,\parallel}(q) = 2\pi k_\text{F}^2 \int\limits_0^\infty dr r J_{0,\chi}(qr) \delta S_{z,\parallel}(r),
\notag
\end{align}
where $J_{l}$ is the Bessel functions of the $l^\text{th}$ order.
In our model we treat the scattering by spin textures perturbatively, 
denoting their sheet density as $\cal N$. 
We assume that the transport scattering rate 
$\tau^{-1} = (2\pi/\hbar) n_i \nu |V_0|^2$ is determined by ordinary impurities with the sheet density $n_i$ ($\nu = M/2\pi \hbar^2$ is the 2D density of states).


The  correlator relevant for the weak-localization problem is 
\begin{equation}
\left[ |V_0|^2 + \left(h_{ex}k_F^{-2}\right)^2|\bm m(\bm q)|^2 \right] \hat{I} - { \hat{V}_{\bm k \bm k'} \otimes \hat{V}_{-\bm k, -\bm k'} }. 
\end{equation} 
This expression should be averaged over the impurity and texture positions
which are assumed to be not correlated.
It follows from Eq.~\eqref{m_par} that 
$m_z$ is invariant while $m_{x,y}$ change their signs under the operation $\bm k, \bm k' \to -\bm k, -\bm k'$.
This differs the scattering by spin textures from both magnetic-impurity and spin-orbit scattering
where $\hat{V}_{-\bm k, -\bm k'}=\hat{V}_{\bm k \bm k'}$.

Introducing the operator of unit angular momentum $\hat{\bm J}$, we obtain that the spin-dephasing rates 
are the eigenvalues of the following operator
\begin{equation}
\label{dephasing}
2\overline{(m_z^2 + m_\parallel^2)} \hat{I} - 2\overline{m_z^2}\hat{J}_z^2 -\overline{m_\parallel^2}(\hat{J}_x^2+\hat{J}_y^2) .
\end{equation}
Here 
$m_{z,\parallel}$ depend on $q=2k_\text{F}\sin{|(\varphi_{\bm k}-\varphi_{\bm k'})/2|}$, and the lines mean averaging over both the initial and final angles $\varphi_{\bm k}$ and $\varphi_{\bm k'}$.
We have taken into account that $m_{x,y}$ are  pure imaginary and $m_z$ is real, see Eqs.~\eqref{m_par}~\footnote{We also use the following relations $\overline{m_im_j} = \overline{m_i^2} \delta_{ij}$,
$\overline{m_{x,y}^2} =-\overline{|m_{x,y}|^2} = - {1\over 2}\overline{m_\parallel^2}$.
}.
%
%
For the singlet ($\hat{J}_i=0$) and triplet states we obtain~\footnote{The spin dephasing rates  $1/\tau_{t0,t1}$ in the triplet channel are calculated by diagonalization of the 3-rank matrix~\eqref{dephasing}. 
Using the relation $\hat{J}_x^2+\hat{J}_y^2=2\hat{I}-\hat{J}_z^2$ valid for the triplet states we get the operator Eq.~\eqref{dephasing}  
in the following form:
$2\overline{m_z^2} \hat{I} + \left (\overline{m_\parallel^2} - 2\overline{m_z^2} \right)\hat{J}_z^2$.
This yields Eq.~\eqref{tau_0_1}.
}
\begin{equation}
\label{tau_0_1}
{1\over\tau_{t0}}= \frac{2}{\tau_{0}}  \overline{m_{z}^2},
\quad
{1\over\tau_{t1}}= \frac{1}{\tau_{0}}  \overline{m_\parallel^2},
\quad
{1\over\tau_s} = {1\over\tau_{t0}}+{2\over\tau_{t1}},
\end{equation}
%
%
where  we introduced the characteristic spin-flip rate $1/\tau_{0}=2\pi  {\cal N} \nu h_{ex}^2 /(\hbar k_F^4)$.
The dephasing times $\tau_{s,t0,t1}$ can be related with the single-particle spin relaxation times 
for spin orientation out of ($\tau_z$) and in the ($\tau_\parallel$) 2D plane by
$1/\tau_z= 2/\tau_{t1}$ 
and 
$1/\tau_\parallel=1/\tau_{t0}+1/\tau_{t1}$. 

The obtained expressions demonstrate that, since $1/\tau_s>1/\tau_{t0,t1}$, the singlet contribution (last term in Eq.~\eqref{d_sigma}) is smaller than that from the triplet channel. Therefore the sign of the magnetoconductivity correction is positive at any size of a spin texture, $\Delta\sigma(B)>0$. However, the form of the magnetoresistance curves 
strongly depends both on the texture size and the Fermi energy, revealing a number of distinct features induced by the chiral character of spin arrangement~\footnote{
For systems with randomly oriented magnetic impurities, all three times are determined by one spin-flip scattering parameter $\overline{m_z^2}=\overline{m_{x,y}^2}$. By contrast, in the studied systems with spin textures, the dephasing rates are different and even can differ by an order of magnitude.
}.


We proceed with considering 
the shape of $\Delta \sigma(B)$ curves for different chiral spin textures $\bm S(\bm r)$. 
Let us mention that for a spin texture with $S_{\parallel} =0 $ (a magnetic impurity with spin directed perpendicular to 2D plane) we have $1/\tau_{t1} = 0$, and $\tau_s = \tau_{t0}$. The correction to the conductivity lacks the dependence on spin potential in this case
and  coincides with a standard curve $\Delta\sigma(B) = (e^2/\pi h) f_2\left({B/ B_\phi }\right)$. 
The change of $\Delta \sigma(B)$ is therefore driven by an appearance of in-plane spin components leading to the spin-flip processes and spin relaxation.

%

To analyze the effect of the chiral spin texture in-plane spin rotation on $\Delta \sigma(B)$ curves 
we provide the numerical calculations for the following texture shape:
\begin{align}
\begin{pmatrix}
\delta S_z(r)
\\
\delta S_{\parallel}(r)
\end{pmatrix} = 
(1-x)^2 \Theta(1-x)
\begin{pmatrix}
\cos{\alpha x}
\\
\sin{\alpha x}
\end{pmatrix} ,
\label{eq_prof1}
\end{align}
where $x=r/a$ with $a$ being the texture radius, $\alpha$ 
controls the in-plane spin inclination, and $\Theta$ is the Heaviside function (we assume that $\delta S_{z,\parallel} = 0$ 
outside a texture core). 
In Fig.~\ref{f1} we plot the dependence of $\tau_{s,t0,t1}^{-1}$ on $k_Fa$ for ${\delta S}_{z,\parallel}$ profiles Eq.~\eqref{eq_prof1} at $\alpha=\pi/2$. 
The increase of a texture radius leads to a more efficient spin-flip scattering and larger $\tau_{t1}^{-1}$ rate. 
When all three times $\tau_{s,t0,t1}$ become different, 
the magnetoconductivity correction 
changes strongly.

\begin{figure}
	\centering
	\begin{minipage}[b]{0.45\textwidth}
		\includegraphics[scale=0.6]{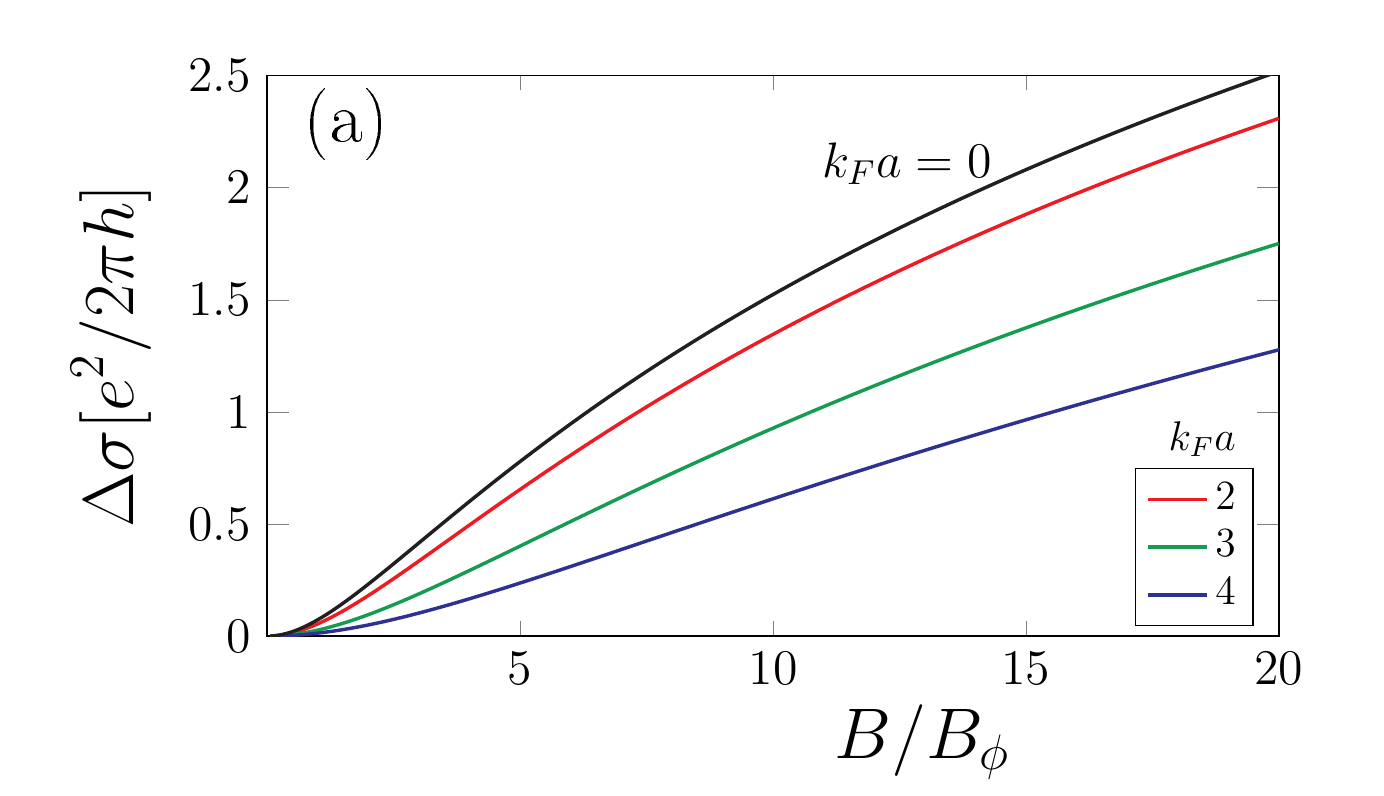}
		\includegraphics[scale=0.6]{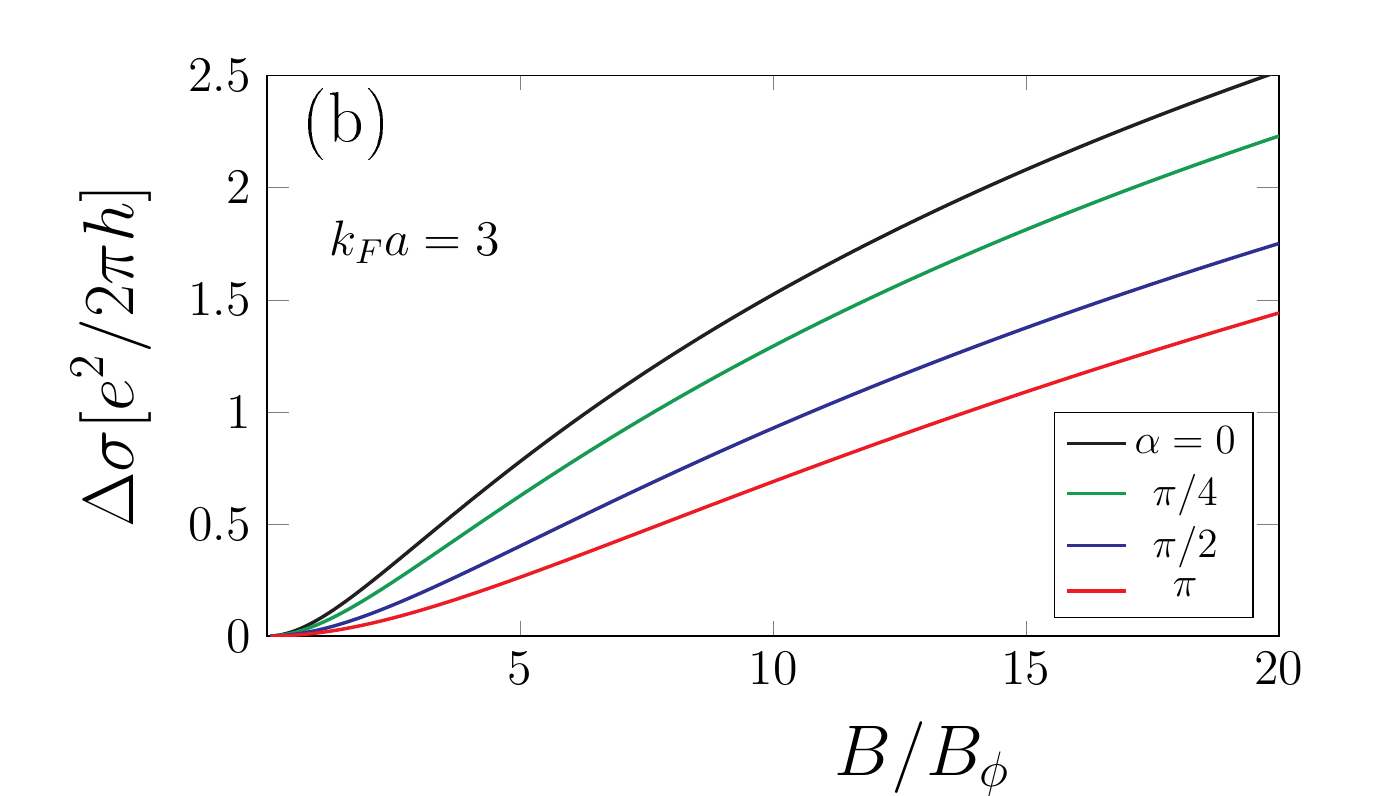}
	\end{minipage}		
	\caption{The magnetic field dependence of the weak-localization conductivity correction 
	at $\tau_\phi/\tau_{0} = 30$
for (a)~different spin texture radius $k_F a$ (at $\alpha=\pi/2$)
and  (b)~different spin inclination angle $\alpha$ (at $k_Fa=3$).
		}
	\label{f2}
\end{figure}

Figure~\ref{f2}~(a) demonstrates $\Delta \sigma(B)$ 
for different texture radius 
suggesting that increase of  the texture size suppresses effectively the conductivity correction.
The inner structure of a spin texture also affects the magnetoresistance. 
The larger is the spin inclination angle and correspondingly the more intense are the spin-flip scattering rates, the more pronounced suppression 
of $\Delta \sigma(B)$ occurs.
In Fig.~\ref{f2}~(b) we plot $\Delta \sigma(B)$ for four different spin configurations 
described by $\alpha = 0, \pi/4, \pi/2$ and $\pi$.
This plot demonstrates the suppression of $\Delta \sigma(B)$ as spin texture is tilted into the plane.



Now we consider the case of large spin splitting assuming that $|k_{\text{F}1}-k_{\text{F}2}| \gg \ell_{1,2}^{-1}$. 
We characterize the spin-flip scattering by the scattering time $\tau_{12}$ which is equal for scattering processes $1\to 2$ and $2\to 1$. 
The spin-flip scattering rate by a skyrmion coincides with the spin relaxation time of triplet state with projection $1$: 
$\tau_{12}=\tau_{t1}$, Eq.~\eqref{tau_0_1}.
In contrast to the coherent subband regime considered previously, for $|k_{\text{F}1}-k_{\text{F}2}| \gg \ell_{1,2}^{-1}$ the spin texture in-plane components affect weak localization only by means of intersubband scattering processes $\tau_{12}$, i.e. without phase factor couplings.
Nevertheless, the change of skyrmion parameters affecting $\tau_{12}$ can still lead to the change of the magnetoresistivity curves. 


Generally, the dephasing times $\tau_{\phi1}, \tau_{\phi 2}$ and the elastic scattering times in the subbands $\tau_{1,2}$ are different due to their dependence on the Fermi wavevectors. 
However, for the parabolic subbands 
the density of states at the Fermi level does not depend on spin index, thus we have $\tau_1 = \tau_2 =\tau$. 
On the contrary, in each 2D subband the dephasing time is linear in the electron density. Therefore we have 
$\tau_{\phi1}, \tau_{\phi 2} = \tau_{\phi}(1\pm \delta) $, where $\tau_{\phi}$ is an average dephasing time, and
${\delta = \Delta/2E_F}$ is the 2D electron gas spin polarization.
As a result, we deal with the  two-subband system with intersubband scattering. The weak-localization conductivity correction for this case is given by~\cite{SSC,FTP,wl_intersubband}
\begin{align}
&\Delta\sigma(B) = {e^2\over 2\pi h} \left[f_2\left({B\over B_+}\right)+f_2\left({B\over B_-}\right)\right],
\\
\nonumber
&B_{\pm} = \frac{B_\phi}{1-\delta^2}\left[ \frac{\tau_{\phi}}{\tau_{12}} + p+1 \pm \sqrt{\left(\frac{\tau_{\phi}}{\tau_{12}} + p\right)^2  + 2p} \right],
\end{align}
where $p=2\delta^2/(1-\delta^2)$
and $B_\phi = \hbar /(2|e| v_\text{F}^2 \tau \tau_{\phi})$ is determined by the average dephasing time $\tau_\phi$ and Fermi velocity $v_\text{F}=\sqrt{2E_\text{F}/M}$.

%
%

\begin{figure}[t]
	\centering	
	\includegraphics[scale=0.6]{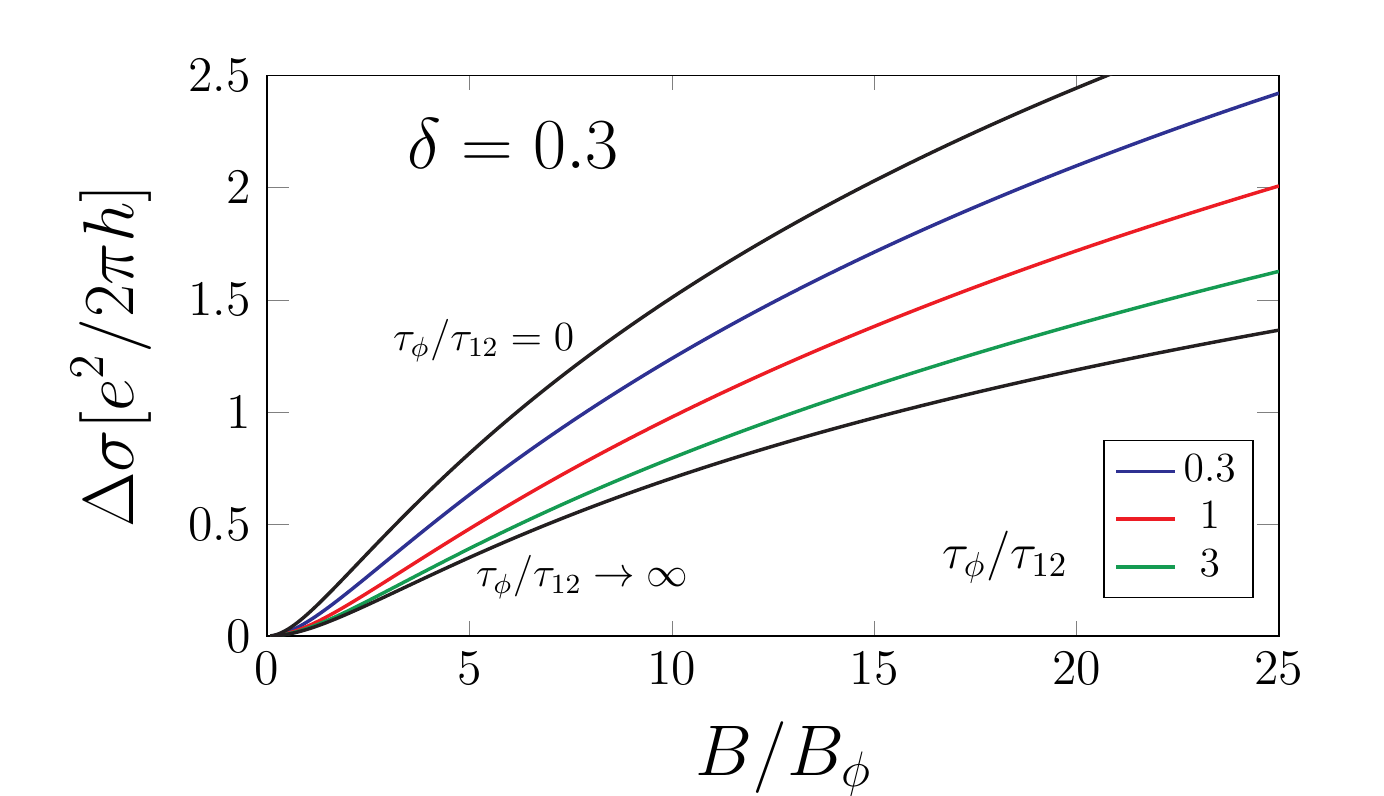}
	\caption{The magnetic field dependence of the conductivity correction 
	for different 
	spin-flip rates $1/\tau_{12}$.
	The 2D electron spin polarization $\delta=0.3$.}
	\label{f3}
\end{figure}

In Fig.~\ref{f3} we demonstrate the evolution of the $\Delta \sigma(B)$-curves driven by the increase of $\tau_{\phi}/\tau_{12}$ for $\delta=0.3$. 
If the intersubband scattering is slow in comparison with the dephasing rates in the subbands 
($\tau_{12}\gg \tau_{\phi}$) then we obtain that spin subbands contribute independently
with $B_\pm=B_\phi/(1 \mp \delta)^2$
corresponding to the 2nd and 1st subbands, respectively.
It is the 2D electron spin polarization $\delta$ that makes $B_{\pm}$ different in this case. 
Increasing the skyrmion size we reduce the intersubband scattering time $\tau_{12}$, thus approaching the opposite limit of fast spin-flips $\tau_{12}\ll \tau_{\phi}$, when we have two distinct characteristic fields 
given by an average dephasing rate and spin-flip rates, respectively:
$B_- =B_\phi/(1-\delta^2)$ and ${B_+ =(2\tau_\phi/\tau_{12})B_- \gg B_-}$.
As a result, the conductivity correction $\Delta\sigma$ is almost twice smaller than in the absence of spin-flip scattering. This is clearly seen from comparison of the curves in Fig.~\ref{f3} corresponding to $\tau_\phi/\tau_{12}=0$ and to $\tau_\phi/\tau_{12}\to\infty$.


\textit{Skyrmion crystals.}
We proceed with considering an important case when skyrmions are spatially arranged in a regular 
lattice. 
We assume the adiabatic electron interaction with skyrmions, i.e. that an electron spin state is everywhere 
co-aligned with a local magnetization 
and the skyrmions do not lead to the 
spin-flip scattering (${\tau_{12} = \infty}$). 
The chiral spatial structure of the magnetization manifests itself 
as a gauge contribution~\cite{BrunoDugaev} to 
the operator of momentum:
${\boldsymbol{p} \rightarrow \boldsymbol{p} \mp (e/c) \boldsymbol{A}(\boldsymbol{r})}$, where the sign corresponds to two electron spin subbands, and 
$\boldsymbol{A}(\boldsymbol{r})$ is determined by a particular spin profile. 
The vector potential $\boldsymbol{A}(\boldsymbol{r})$ 
leads to the 
``topological'' magnetic field acting on electron orbital motion. Its averaged component perpendicular to 2D plane is given by:
\begin{equation}
{B}_T = \pm \frac{\phi_0}{4\pi} \langle \boldsymbol{S} \cdot \left[ \partial_x \boldsymbol{S} \times \partial_y \boldsymbol{S} \right] \rangle \equiv \pm Q {\phi_0 \over \mathcal{A}}, 
\end{equation}
where $\phi_0 = h c/|e|$ is a magnetic flux quantum, $Q$ is an integer number, and $\mathcal{A}$ is an area of the crystal unit cell. 
Let us note that the sign of the emerging magnetic field 
is opposite for two electron subbands, 
which modifies the 
anomalous magnetoresistance in a nontrivial way. 

\begin{figure}[t]
	\centering	
	\includegraphics[scale=0.6]{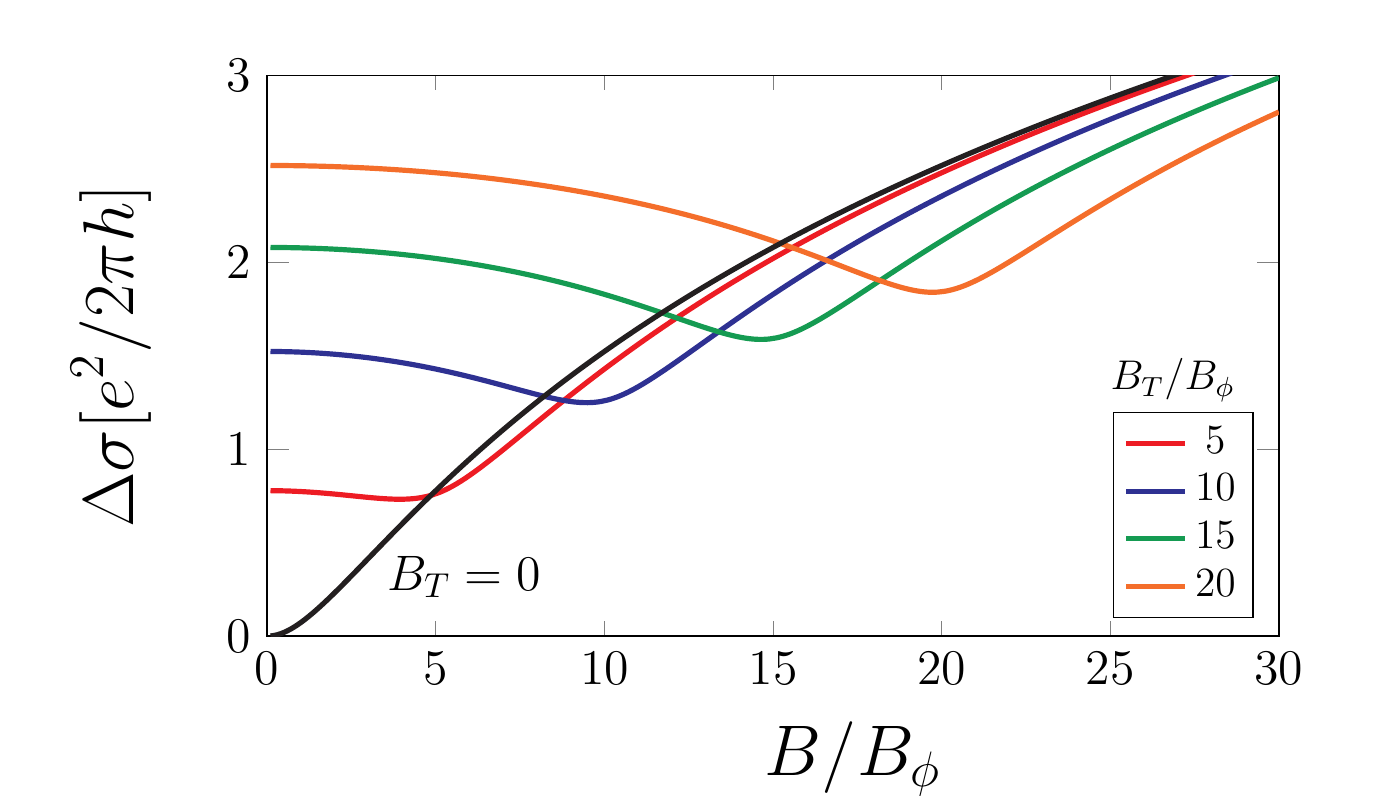}
	\caption{The dependence of $\Delta \sigma(B)$ for skyrmion crystals with different values of the topological magnetic field 
	${B}_T$.}
	\label{f4}
\end{figure}

It is worth noting that the assumption of the spin adiabaticity  ($\tau_{12}=\infty$) is typical for a strong exchange interaction when the condition $|k_{\text{F}1}-k_{\text{F}2}| \gg \ell_{1,2}^{-1}$ is additionally fulfilled. At that 
the electrons from different subbands 
contribute to the conductivity independently, with the corresponding total 
magnetic field being a superpositon of the external field $B$ and the topological one ${B}_T$:
\begin{equation}
\label{skyrm}
\Delta\sigma(B) = {e^2\over 2\pi h	} \left[f_2\left({B + {B}_T\over B_\phi}\right)+f_2\left({B - {B}_T\over B_\phi}\right)\right].
\end{equation}
Here we assume equal dephasing times in the spin subbands.

Figure~\ref{f4} shows the dependence of $\Delta \sigma(B)$ on the external magnetic field $B$ for different ratio ${B}_T/B_{\phi}$. 
This dependence is featured by the emergence of a noticeable dip starting from $B_T/B_\phi \approx 4$. 
The position of this minimum
is unambiguously associated with the magnitude of the topological field ${B}_T$. 

We note that 
the other transport phenomenon 
takes place in skyrmion crystals, namely the topological Hall effect. 
Therefore we argue that 
when the topological Hall effect is experimentally observed, 
the longitudinal conductivity would also experience a modification 
due to weak localization according to Eq.~\eqref{skyrm}. 

In summary, we  developed the weak-localization theory for 2D electron systems with chiral spin textures. We demonstrated that, in  disordered arrays of spin textures,  the weak localization  is featured by a specific behavior that cannot be attributed to any other mechanism. For skyrmion crystals, the sign-alternating magnetoresistance is predicted.

Financial support of the Russian Science Foundation (Project No. 17-12-01265) and
Foundation for advancement of theoretical physics and mathematics ``BASIS'' is acknowledged.

\bibliography{Skyrmion}

\end{document}